\documentclass[12pt]{article}
\usepackage[left=2cm,top=2cm,right=2cm,bottom=2cm,nohead,nofoot]{geometry}
\usepackage[pdftex]{graphicx}
\usepackage{setspace}
\usepackage[square]{natbib}
\usepackage{amssymb}
\usepackage{verbatim}
\usepackage[usenames,dvipsnames]{color}

\newcommand{\imloc}{}
\newcommand{\baseloc}{../../../}

\begin{document}

\title{Variance Decomposition and Replication In Scrabble: When You Can Blame Your Tiles?}
\author{Andrew C. Thomas}
\date{\today}

\maketitle

\begin{abstract}
In the game of Scrabble, letter tiles are drawn uniformly at random from a bag. The variability of possible draws as the game progresses is a source of variation that makes it more likely for an inferior player to win a head-to-head match against a superior player, and more difficult to determine the true ability of a player in a tournament or contest. I propose a new format for drawing tiles in a two-player game that allows for the same tile pattern (though not the same board) to be replicated over multiple matches, so that a player's result can be better compared against others, yet is indistinguishable from the bag-based draw within a game. A large number of simulations conducted with Scrabble software shows that the variance from the tile order in this scheme accounts for as much variance as the different patterns of letters on the board as the game progresses. I use these simulations as well as the experimental design to show how much various tiles are able to affect player scores depending on their placement in the tile seeding.
\end{abstract}


\doublespacing

\section{Introduction}

In the game of Scrabble, there are at least three sources of variation in score: the ability of the players, the order in which tiles are drawn from the bag, and the pattern made by the tiles on the board as the game progresses. Randomness from the bag and the board makes it more difficult to tell if one player is better than another; the more variation there is, the easier it is for an inferior player to win a head-to-head match against a superior player, and the more matches it would take to figure out the true ability levels for a set of players. Reducing uncontrolled variability is a classic problem of experimental design, so surely there is something that can be done to address this without necessarily compromising the original game.

Like many in the mathematical sciences, I've been a player and fan of the game of Scrabble since childhood.  My own personal fascination with the game to this day comes from the tension between its two main groups of fans: literary types tend to enjoy playing creative and interesting words, and quantitative types often memorize reams of words purely for their use in the game without regard to their meaning. (I fall into either camp, typically depending on whom I play against.) 

Far from being a pure game of skill, luck and chance play a significant role in the way a game can develop. Each player has (at most) 7 tiles on their rack at any one time, replenished from a bag containing those tiles that remain from the 100 at the beginning of the game; the player can also choose to swap a number of tiles with replacements from the bag. And to top it all off, every move affects every subsequent move, both in the tiles that remain in play and on the configuration of the board once those words are played. One reason that the letter S is considered valuable is that it can instantly pluralize many English nouns, providing a prime opportunity to ``hook'' a seven-letter word onto an existing word for extra points.

High-level games place considerably more emphasis on plays where all seven of a player's tiles are used; these ``bingos'' score an additional 50 points on top of the word value. This has at least two major consequences to the way a game will unfold. First, the more letters that are played, the more potential spaces are open on the board for other plays, including more bingos, so that scores can increase more rapidly for both players. Second, the incentive to create words of seven letters or longer gives additional value to more frequently drawn tiles, and especially to the two blank tiles that can substitute for any letter; even though they have no direct value to the player, their indirect value in producing bingos is said to make them the most valuable tile in the bag.

As a player of the game, I would love to remove as much luck from the game as possible to get a better estimate of my own skill level against that of others, and in cases where both the blanks are drawn by one player, there is certainly a feeling that on this scale, randomness is a curse rather than a blessing; as a practicing statistician, I want to do it as efficiently as possible, getting a better gauge of ability from fewer games played, especially when there is money on the line at a tournament where players are grouped by their estimated skill level.

\section{Introducing The Two-Sided Draw Method}

The principle is to give each player as close to the same tiles drawn if the match were repeated, yet still preserving the outward appearance of randomness to the two players involved. The notion is that if many different pairs of players are given the same tile order, the only remaining variation will come from the board and the player's own abilities, not the order in which tiles are removed from the bag, so that a player can be compared both against their opponent across the board but also their peers with the same potential tile selection. This would give the option of a tournament option similar to duplicate Contract Bridge that still features the adversarial nature of traditional tournament Scrabble.\footnote{``Duplicate Scrabble'' is \href{http://en.wikipedia.org/wiki/Duplicate_Scrabble}{already the name of a different variant of the game}, common in Europe in which players are given a board position and seven tiles and challenged to find the best play. The game has no defensive component to it and so is fundamentally different from the strategy in two-player games.}

Additionally, this set-up allows us to conduct simulations that better gauge the value of a tile in the context of the game with a simple two-level structure: many tile settings can be produced, with each setting replicated a large number of times. The position of a tile within the overall structure will be associated with the end score of one player, and the score difference of the two, giving a meaningful way of quantifying a tile's value.

\begin{figure}
\includegraphics[width=\linewidth]{\imloc 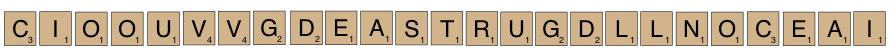}
\caption{The reserve tiles are placed in a predetermined order, unknown to the players.\label{t1}}
\vskip 2cm
\includegraphics[width=\linewidth]{\imloc 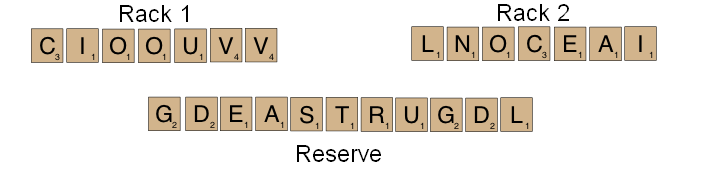}
\caption{Each player draws tiles off their own end of the reserve sequence. When repeating this tile order, each new player in these positions will receive many of the same tiles, depending on the number they play and each player's discards.\label{t2}}
\vskip 2cm
\includegraphics[width=\linewidth]{\imloc 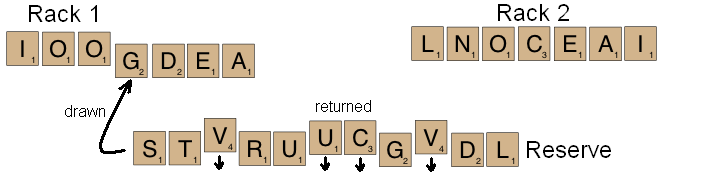}
\caption{When exchanging tiles, the new draws are first taken from the player's drawing position. The discarded tiles are inserted uniformly at random within the reserve sequence.\label{t3}}
\end{figure}

Figures \ref{t1}, \ref{t2} and \ref{t3} demonstrate the mechanism for ensuring that Player 1 will tend to receive the same tiles in the same order if the game were repeated, and likewise for Player 2. First, the tiles are placed in a predetermined order (as seen in Figure \ref{t1}) that is invisible to the players. When Player 1 replenishes their rack, they draw tiles from the front of the order; Player 2 draws from the back. This way, even if the players were to play words of differing lengths in different replications, they would be just as likely to receive the same tiles. As the game progresses, tiles are removed from each end of the sequence until there are no more to draw from. 

A player always has the option of exchanging some or all of their tiles in lieu of playing a word on the board. If this is the case, the letters can be placed uniformly at random throughout the remaining sequence, so that when they would be redrawn would still be invisible to the players of the game.\footnote{Technically, it is possible to predetermine where any tile combination would be distributed among any remaining tile sequence before the game was played, as a way of further reducing the variance between replications of games. However, this seems even to me like overkill, given the combinatorial size of the problem and the minimal gain that would likely be obtained from this.}

This initial sequence of letters can then be used for all games. At present, this is technically infeasible to do manually, since it would require the design of an apparatus for holding tiles in order without being seen by either player, as well as a method of redistributing exchanged tiles without either player being able to track it. It is, however, ideal for inclusion in computer-based Scrabble games, where the physical aspects of the problem are no longer in place. This also gives us the benefit of being able to simulate a very large number of games to get some sense of how the method might work if deployed for real

\subsection{Testing The Method with Scrabble AI}

There is an abundance of software that can duplicate the Scrabble experience for human players, including online services like Scrabble for Facebook and the international site \texttt{isc.ro}. When it comes to publicly available computer players for Scrabble, there are at least two academic projects that have been developed, published and tested: Maven \citep{sheppard2002ws} was among the first publicly released and tested program to compete against, and defeat, championship-level Scrabble players. \href{http://www.quackle.org}{Quackle} is another, first released in 2006, that offers several different levels of difficulty for computer players, along with a pleasing interface and computer suggestions for human player moves.
Quackle was the best choice for running this test due to its open source nature and its infrastructure: the software package includes a ``test harness'' for examining the effects of various changes in the AI, as well as for simulating many games in sequence. I subsequently adapted the C++ code to use the two-sided draw method and take as input any given tile sequence and ran the interface from a subroutine written in R.


For each game, I set two Quackle ``Speedy Player'' computer players (henceforth known as ``bots'')  against each other. This particular AI evaluates potential moves without any active forward looking, calculating only the short-term ``utility'' of a move: the value of the played word plus a pre-computed ``leave'' value, or the estimated value of the remaining tiles in combination with each other, plus a small adjustment for the number and quality of locations that are now accessible to the opponent. For example, a leave with two Us is significantly less valuable that one with two Es, based on the number of potential words that can be formed with these letters (especially bingos). The play with the highest utility is chosen. While \citet{richards2007oms} remind us that modelling the opponent's likely strategy is also important to the forward projection problem, the Quackle Speedy Player is shown to be a competent player without this addition. However, because the Quackle Speedy Player only seeks to maximize its own score, without regard to defensive positioning, it would be improper to conclude that the valuations made by the AI, and subsequent estimates, would correspond directly to the decisions made by expert human players; it does the job wonderfully for the sake of proving and testing the two-sided draw.

Normally, the Quackle Speedy Player bot uses a deterministic method to select a move, so that if two of these players faced off against each other a number of times with the same tile order, the exact same game would result every time. To account for this, I adjust the move selection process by adding a Uniform$(-1,1)$ random variable to the utility of each potential move calculated. While that there would be some probability of choosing a slightly suboptimal move, including one of a number of permutations or placements with the same score, there would be zero chance of selecting a word that was markedly below the maximum utility (at least two points below would be impossible.) This small perturbation is shown in simulation to be both necessary for exploring the real game, and sufficient to introduce a great enough variety in the outcomes of games due to the board while not impairing the AI.

\section{Comparing Scrabble against Words With Friends}

Of the many imitation versions of the game of Scrabble that are available online, the most popular is Words With Friends, created by Zynga, the company likely best known for the Facebook game FarmVille. The principle of the game is the same, though there are many differences (summarized in Table \ref{table:comparisons}). Among the changes made are a different board design, a change in tile distribution and value, and a lesser bonus for a ``bingo'' play. (A nice primer on some strategic differences can be found \href{http://www.robotsfighting.com/?p=60}{here.}) The two-sided draw and Quackle software can be used for either version of the game once the appropriate parameters are loaded.

\begin{table}
\begin{center}
\begin{tabular}{c|c|c}
\hline
\hline
Property & Scrabble & Words With Friends \\
\hline
 & \includegraphics[width=0.37\linewidth]{\imloc 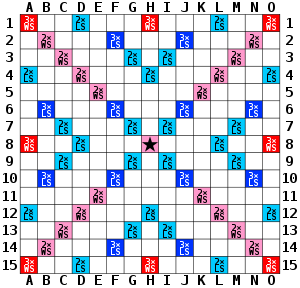} & \includegraphics[width=0.37\linewidth]{\imloc 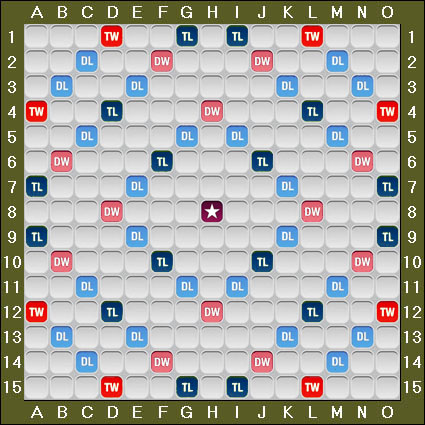} \\
\hline
Layout & Radial & Concentric \\
Tiles & 100 & 104 \\
Number of S'es & 4 & 5 \\
``Bingo'' bonus & 50 & 35 \\
\hline
\end{tabular}
\end{center}
\caption{Differences between the standard Scrabble and Words With Friends configurations. The player to play first must play at least one tile on the center square; this is a Double Word Score bonus in Scrabble but not in Words With Friends.\label{table:comparisons}}
\end{table}

49,400 different tile orders were generated for Scrabble, as well as 43,800 for Words With Friends. For each order, 100 games were played between two Quackle Speedy Players for a total of 4,940,000 and 4,380,000 matches each. Results are first collected and summarized for each tile order; these summaries are then used to compare different tile orders. Both the total score for one player and the difference in scores between each are of interest, though only the latter -- whether one player had more or fewer points than their opponent -- determines the winner in tournament play. In almost all cases, each player had access to at least 40 tiles. 

Player 1 had an average score of 435 points in Scrabble and 464 points in Words With Friends. While the latter game has a smaller ``bingo'' bonus, an increased total number of tiles and changes in value are factors that can increase the mean score; also new is the possibility of a ``triple-triple'', in which a letter can have nine times the value due to a combination of a triple-letter and a triple-word bonus.

There are several outcome quantities that can be obtained for each game other than the final score, including the specific tiles that each player used as well as the total number available to each player. Each of these is technically an intermediate outcome on the way to the final game score, so figuring out any truly causal questions (``if Player 1 played the Q, what would their difference in score be?'') is slightly trickier. It is much cleaner to start from the placement of each letter in the initial sequence and associate that with the final score in order to get the value of a letter, especially since the players would have the option of exchanging their letters with new ones from the bag. 

\subsection{Total Variance, From The Bag and On The Board}

\begin{figure}
\begin{center}
\includegraphics[width=0.48\linewidth]{\imloc 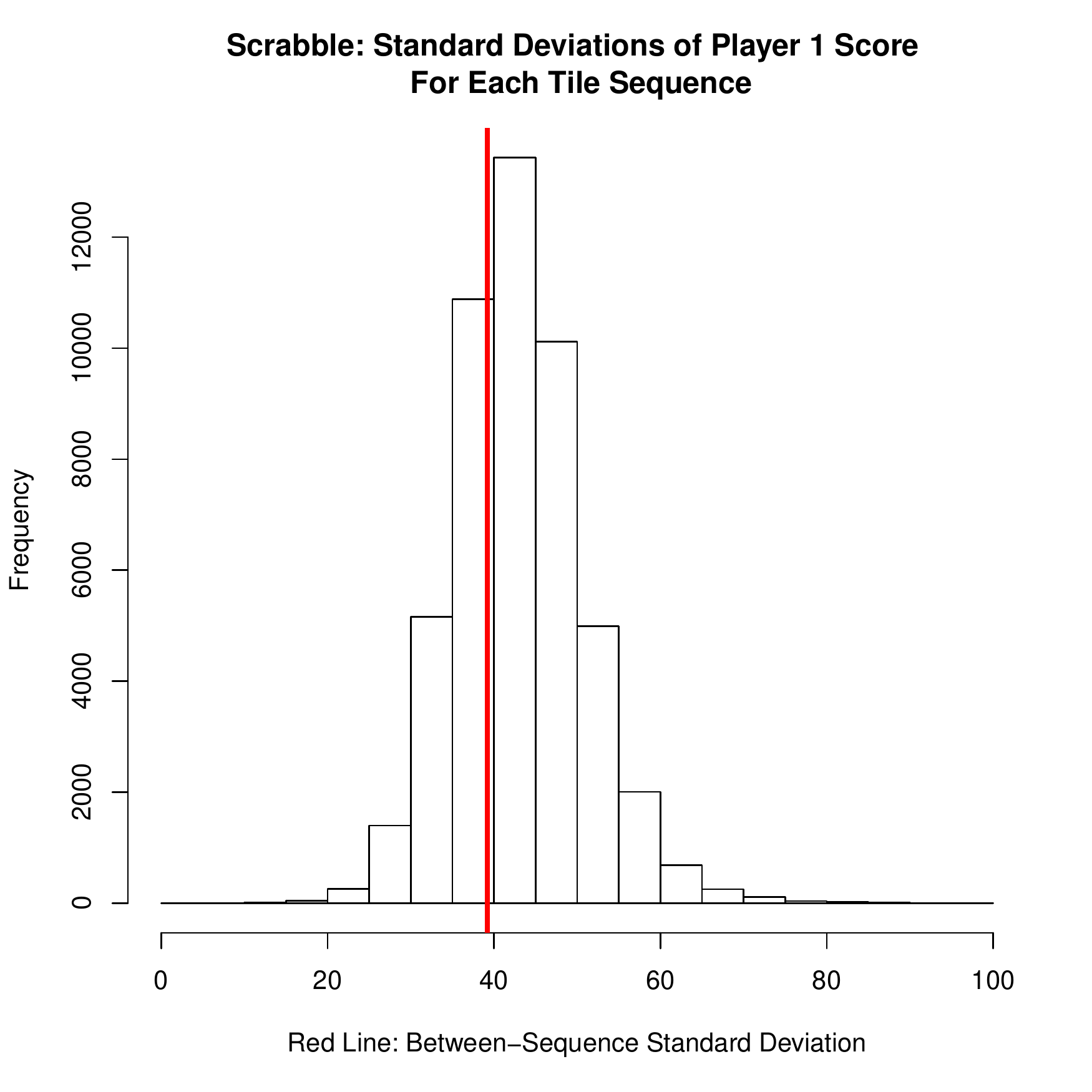}\includegraphics[width=0.48\linewidth]{\imloc 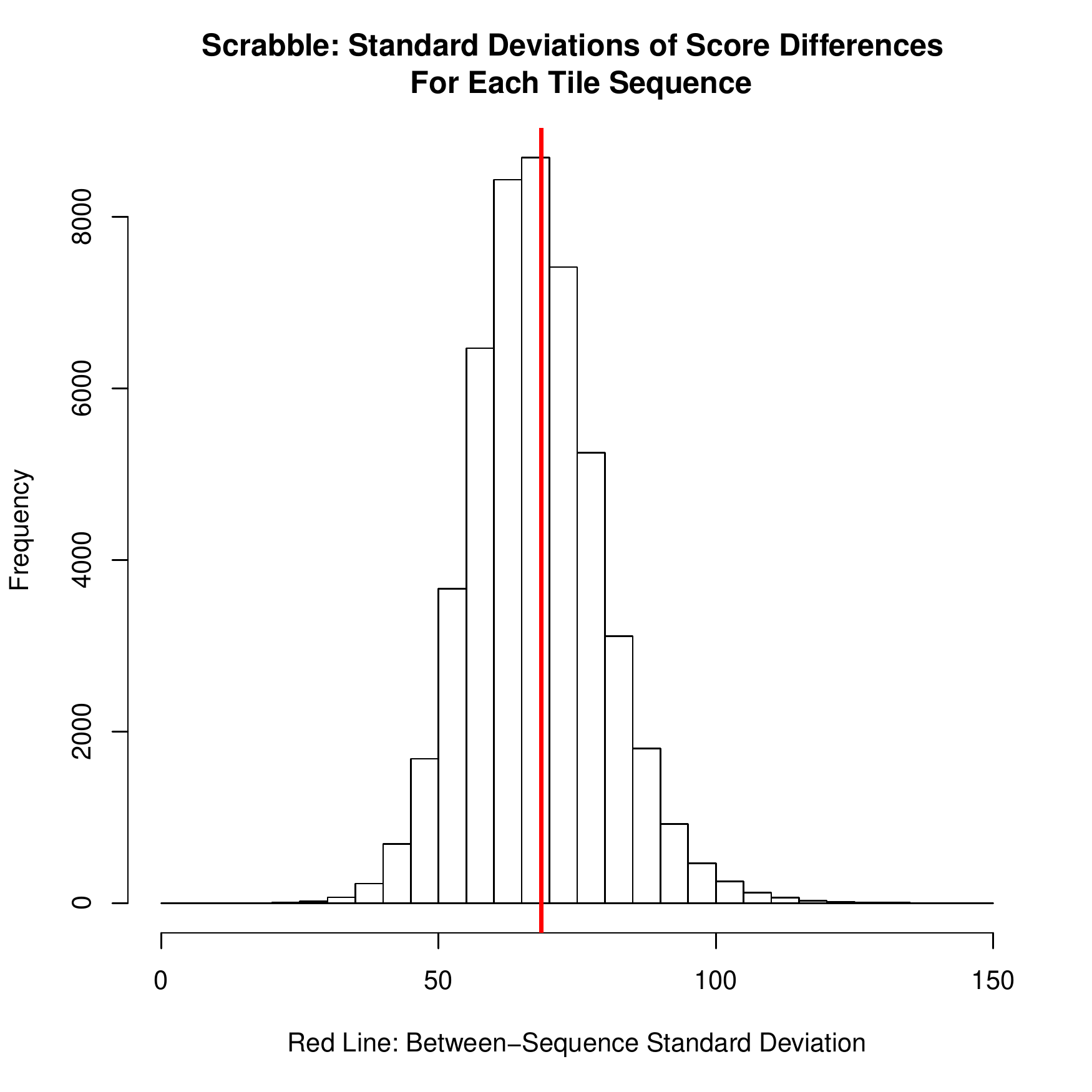}
\includegraphics[width=0.48\linewidth]{\imloc 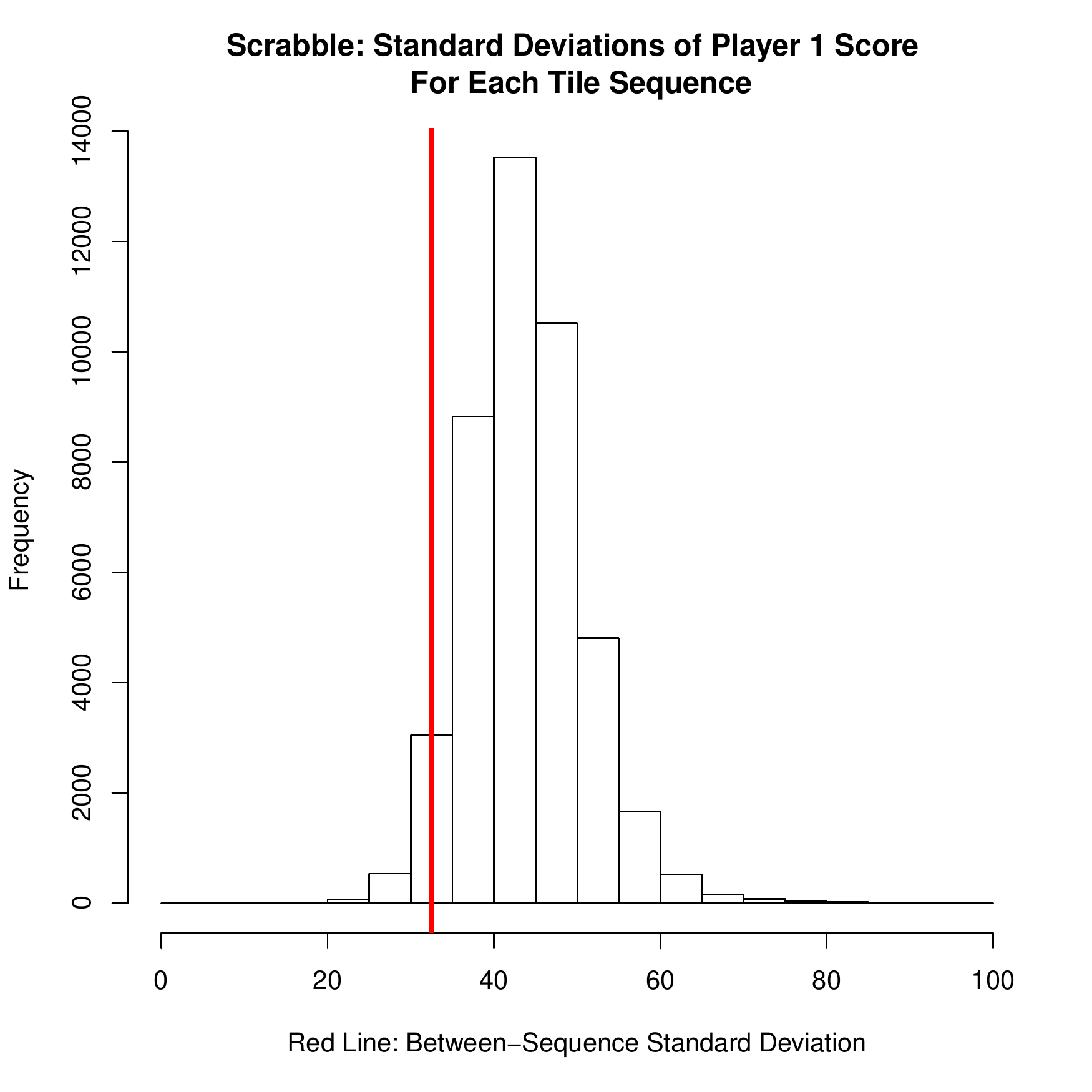}\includegraphics[width=0.48\linewidth]{\imloc 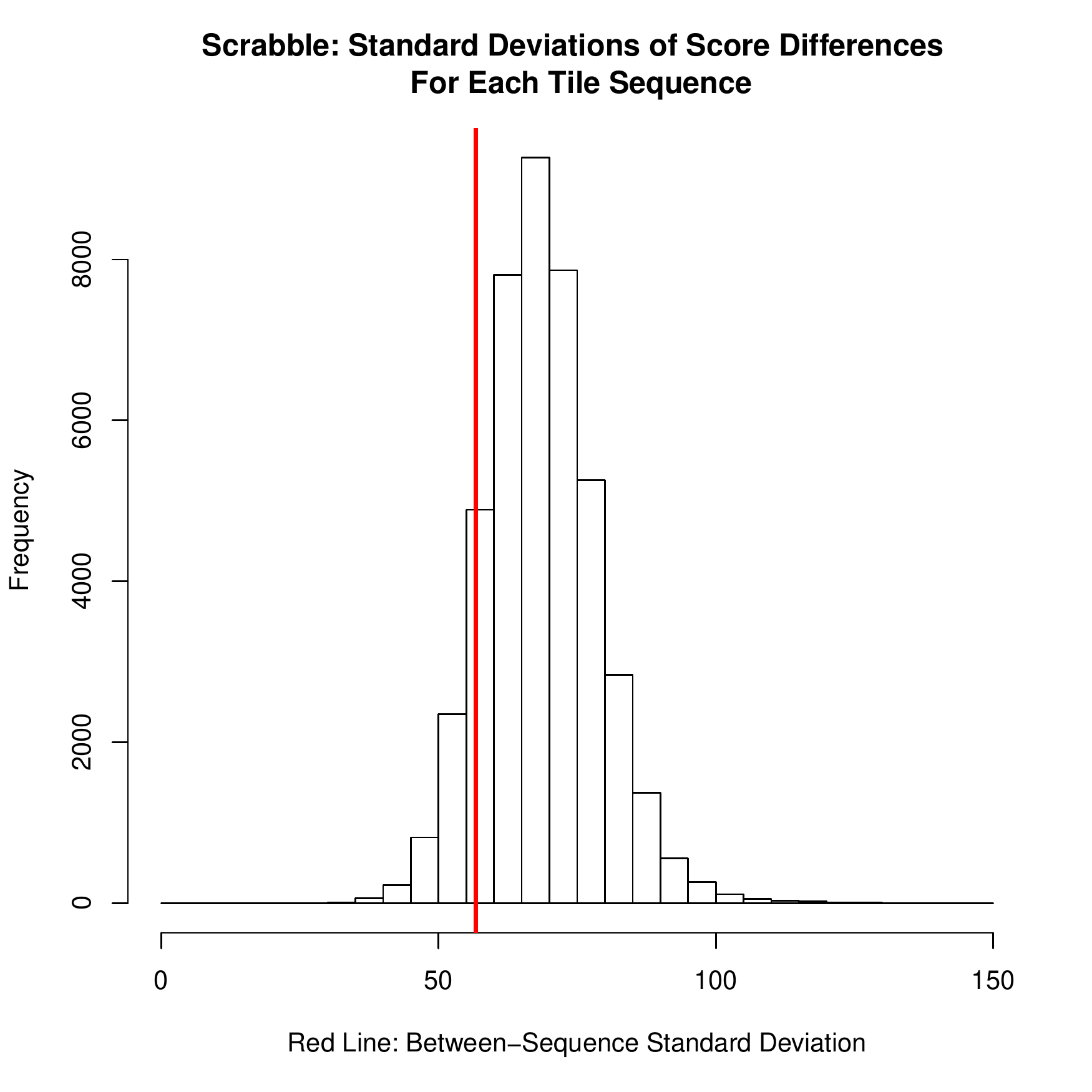}
\end{center}
\caption{Comparing the standard deviations for Player 1 score and the difference in scores, between and within tile sequences, for Scrabble and Words With Friends. The average between-sequence deviation is seven points smaller for Words With Friends than for Scrabble, likely due to the reduction in bingo bonus. The within-sequence deviation, however, is slightly greater for Words With Friends, suggesting that the reduced bingo bonus and changed tile values have little bearing on how a match would have played out given that the draw order was preset. \label{fig:histograms}}
\end{figure}

For any one tile order, the mean and variance of the score for one player, and of the difference in scores between the two players, are calculated. As shown in Figure \ref{fig:histograms}, there is a wide range of score and score difference variability across the various simulated tile orderings. 

The red line in each plot represents the variance of the mean values, and represents the variability between different tile orders. For the mean of player 1's score, the bag represents 44\% of the score variance in Scrabble and 34\% in Words With Friends; this jumps to 50\% and 40\% for the difference between scores for each player. This very substantial proportion of the variance could be reduced for live games if many pairs of players had access to the same tile order.


\subsection{Between Bots, Whoever Goes First Has An Edge}

Taking the mean score of the first player and subtracting the mean score of the second, Player 1 is shown to have a net lead of roughly 14 points per game over their opponent in Scrabble. There is an indisputable bonus to going first in this case. The size of the effect is small compared to the 100 point standard deviation across all tile orders, but may present a sizeable bias in those cases where the within-order standard deviation of score difference is small; as it is 60 or less in 25\% of simulations, there is ample reason to consider a modification to the rules to remove this effect.

One of the features of the Scrabble board is the presence of ``premium'' tiles, for which a letter or word score is doubled or tripled. One feature of the board is that since the player who goes first has no tiles on which to build their words, their first play receives a double score. It may be time to consider a tournament board where this bonus is removed, or at least adjusted so that this advantage is nullified. Interestingly, Words With Friends does have a change of this type, but still retains a first-to-play bonus of roughly 10 points. This is likely because a double word score is still accessible to any five-letter play for the opening turn (as shown in Table \ref{table:comparisons}.)

\subsection{Blanks Are Worth An Additional 25-30 Points Each, With Some Exceptions}

\begin{figure}
\begin{center}
\includegraphics[width=0.9\linewidth]{\imloc 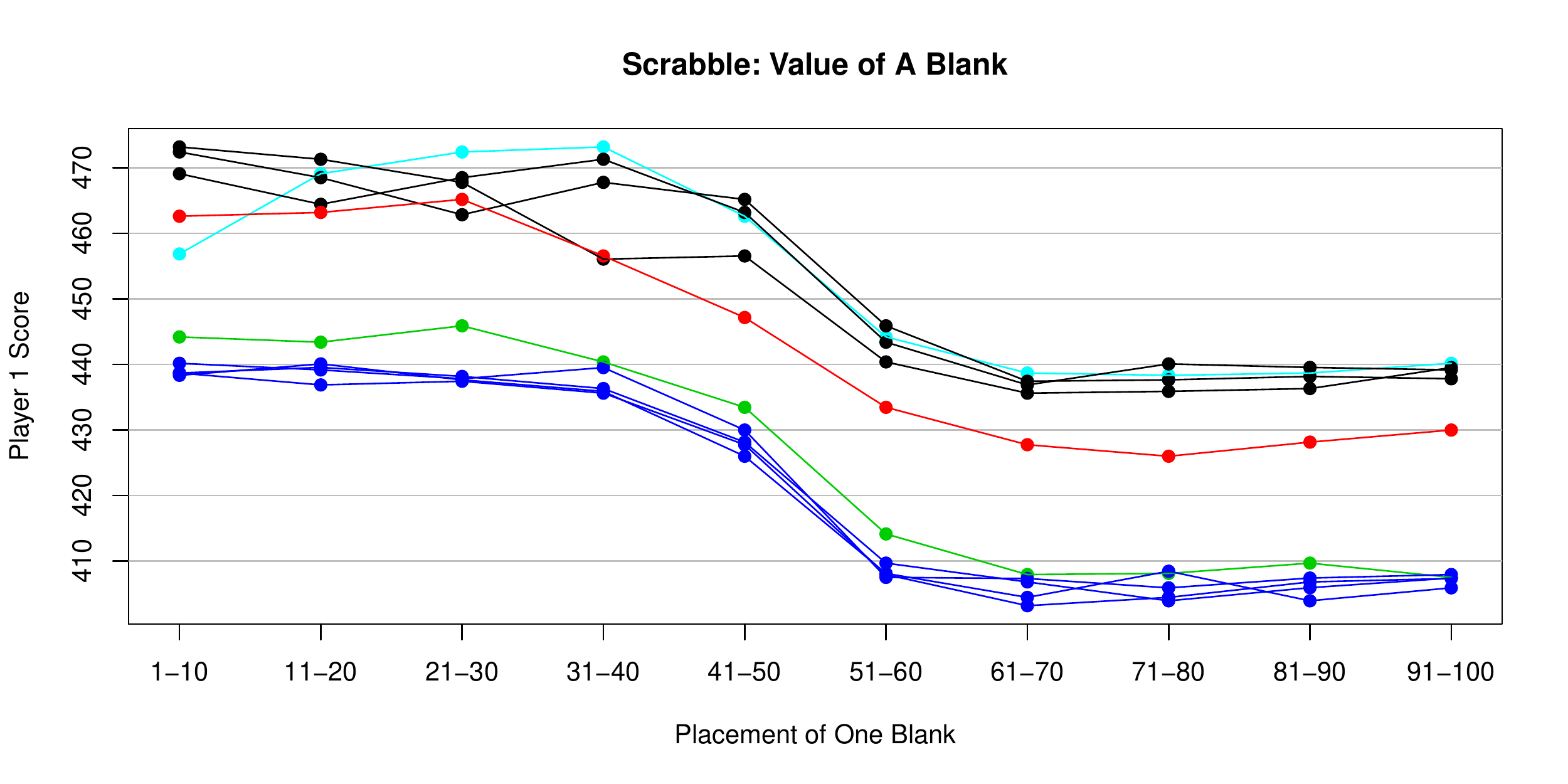}
\includegraphics[width=0.9\linewidth]{\imloc 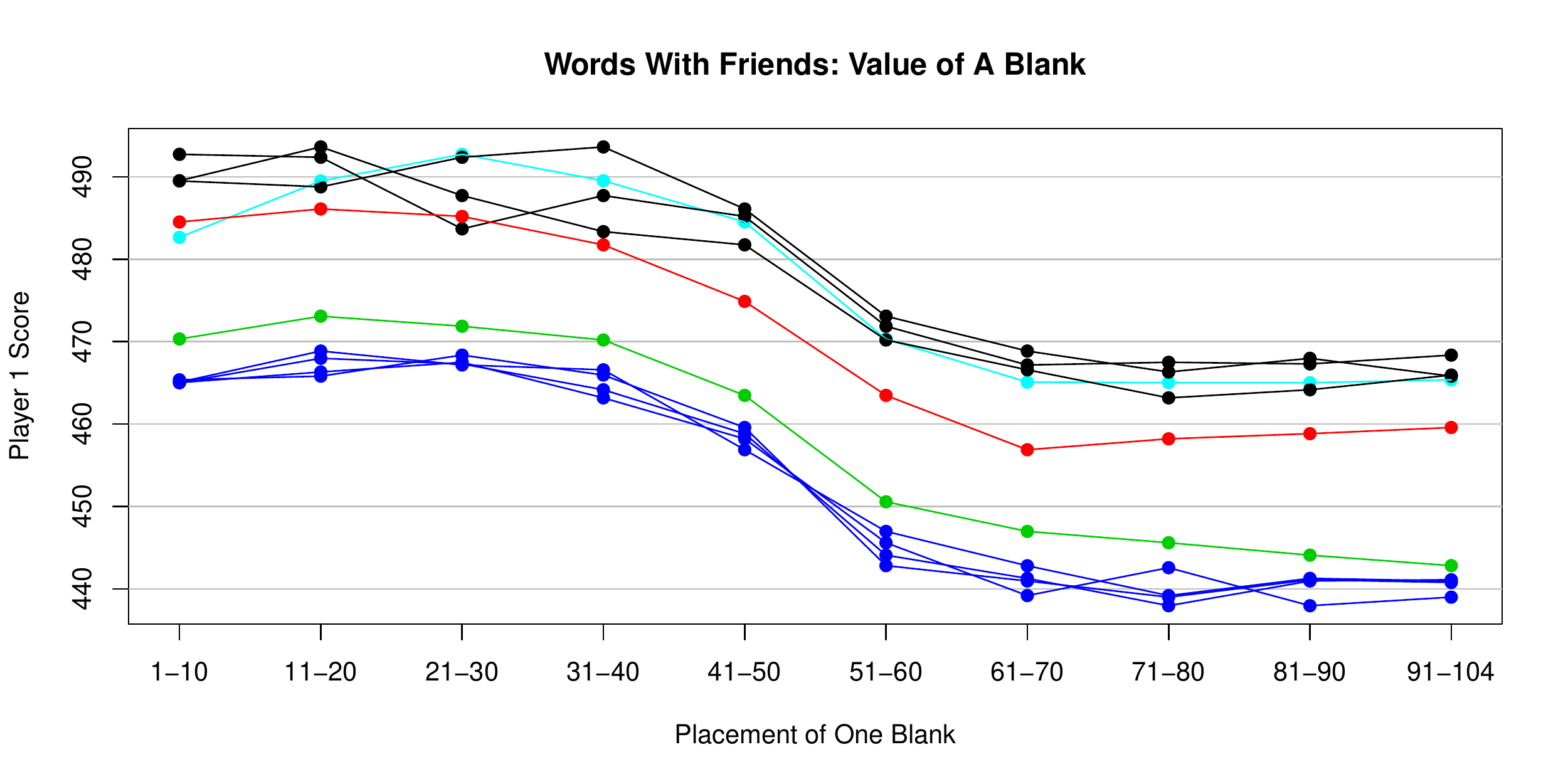}
\end{center}
\caption{Average game score for player 1 given the location of each of the blanks. For Scrabble, the average difference in score from the top and bottom groups, representing a blank in locations 1-50 and 51-100 (or 104), is roughly 30 points; for Words With Friends, the difference is about 25 points. \label{fig:blank-value}}
\end{figure}

Since there are two blanks in the tile set, assessing their value for each player depends not only on the location of each, but also their location relative to each other. Grouping the placement of each blank within 10-tile groups, we have a sense of the value of the blank given the chance that one player will discover it, and at what point during the game this will occur. In Figure \ref{fig:blank-value}, each line represents the placement of one blank in a decile, and each point on the line represents the position of the other blank.

The green and blue lines represent the score for the first player if one of the blanks is located in the latter half of the order, so that the first player is unlikely to ever draw it; if they did, it would be at the end of the game when the chance of scoring a bingo is minimal. In these cases, the difference is clear: if the other blank is in the first 40 tiles, the player scores about 30 more points than if it were in the latter half (25 in Words With Friends), and if in tiles 41-50, the difference is about 15 points (12 in WWF), so that the presence of the blank near the endgame is not as beneficial as it would have been earlier. 

This pattern is also reflected in the red line, where one blank is in the fifth decile, as it is typically worth half the value of a blank earlier on. The exceptions are when other blank is likely to be obtained by the first player, since under this case, the player is far more likely to have played a prior bingo and therefore played more tiles; in this case, there are still more tiles in the bag when the second blank would be drawn, and it would retain more of its value being played sooner than the end game.

The cyan line represents the score for the first player if one of the blanks lies in the first 10 tiles; the black lines are for the second, third and fourth deciles. While the 30-point rule holds in most cases, if both blanks are in the first 10 tiles their combined value is significantly diminished. This is likely due to the fact that if both blanks appear on the same rack, there is only one potential bingo play rather than the two that would be expected if they were separated.

\subsection{How Much Is Each Tile Effectively Worth?}


The same method to estimate the value of a blank can be used to find the value of any other tile in the bag, though for those with more than one copy it is considerably trickier to deal with the issue of a change in location. It is worth going through this with the S, due to its importance as an extender of many English words. Once again, the grouping and position of each of the S'es plays a role in their value to each player. 

\begin{figure}
\begin{center}
\includegraphics[width=0.9\linewidth]{\imloc 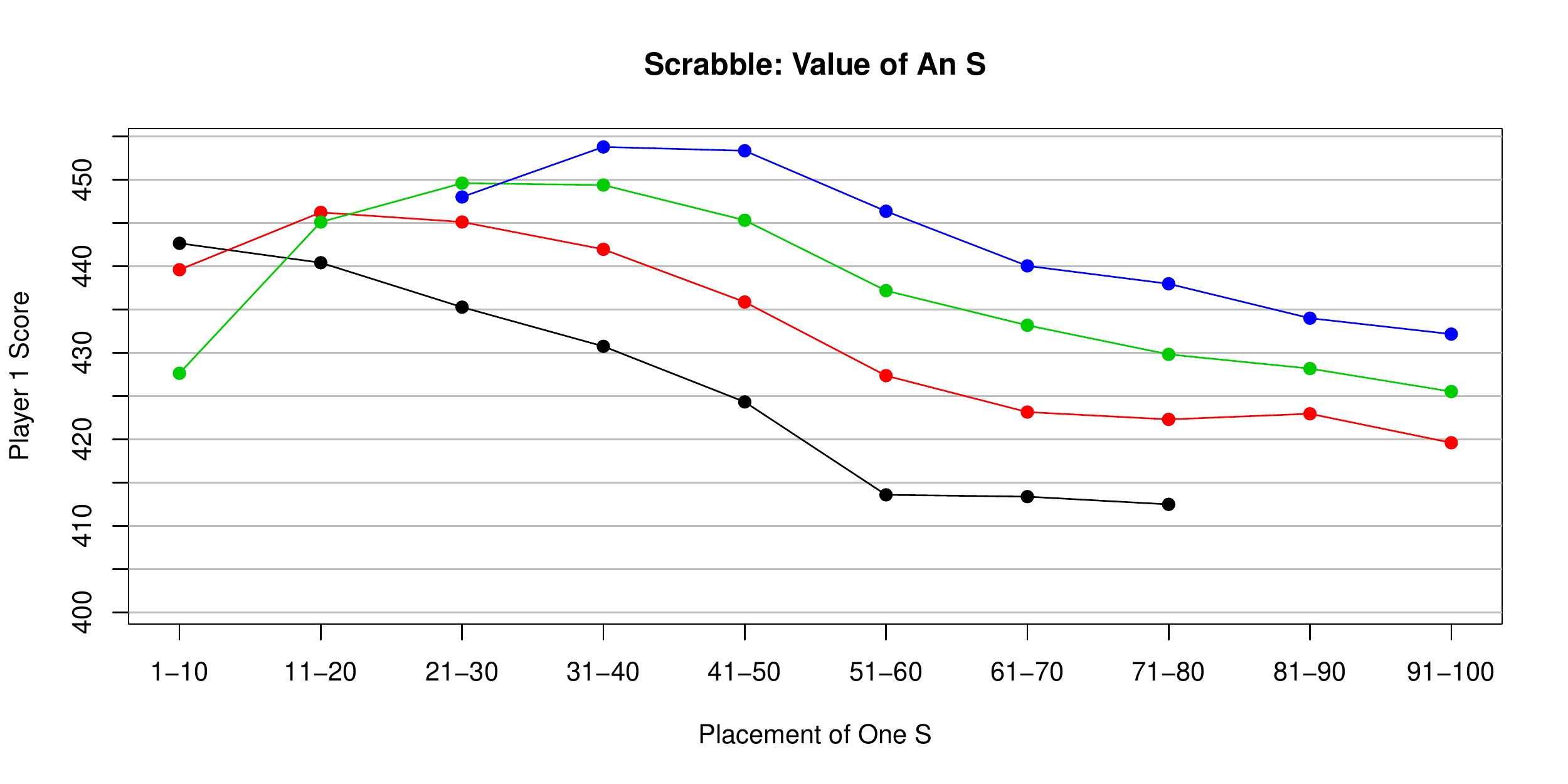}
\includegraphics[width=0.9\linewidth]{\imloc 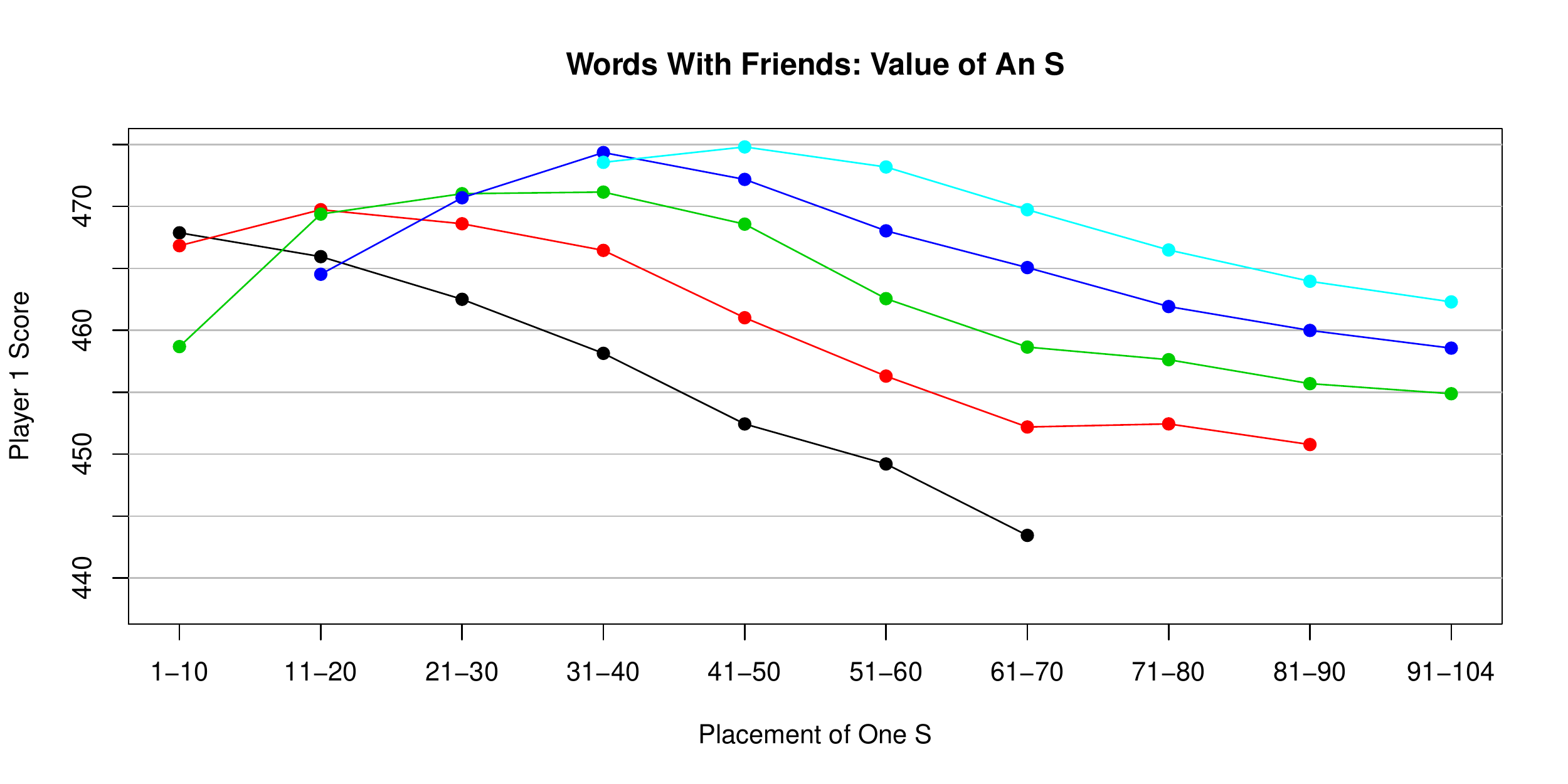}
\end{center}
\caption{Average player 1 score for the location of each S in the tile sequence. The black, red, green and blue lines represent the mean scores for the first player if the first, second, third or fourth S falls in each decile; for example, the blue dot under ``41-50'' means that for that subset of games, all four S'es were in the front half of the tile order. A dot is present if there are at least 10,000 simulated games in that group. The difference between two dots in the same decile therefore represents the additional value of more S'es before that point. In particular, the columns for ``31-40'' and ``41-50'' suggest the value for having each additional S available to player 1, and that the net increase in score is roughly 10 points per S in Scrabble, and 7 points in Words With Friends, given that there is a good deal of space available between each of them. The consequence of the S'es being closer together can be seen in the cases where the first player has many S'es at the beginning of the order, in which the differences in score are considerably reduced. Indeed, in the few simulations where all four S'es were in the first 20 tiles, the mean score was less than in those cases where at least one S came later.  \label{fig:s-value}}
\end{figure}

Figure \ref{fig:s-value} shows the mean score of a game if we know a particular S is in that block. The blue line, for example represents the fourth S; the block where this is in location 31-40 implies that three more Ses must lie ahead of it, and are almost certain to be picked up by Player 1. Dropping down to the green line means that on average we have moved one S from before this position to after it, and that this is the value of an S relative to a tile drawn from the bag at random. From this scheme, roughly 10 points separate the lines for Scrabble, and 7 points for Words With Friends; this discrepancy can be explained largely by the change in the ``bingo'' bonus.

\begin{figure}
\includegraphics[width=\linewidth]{\imloc 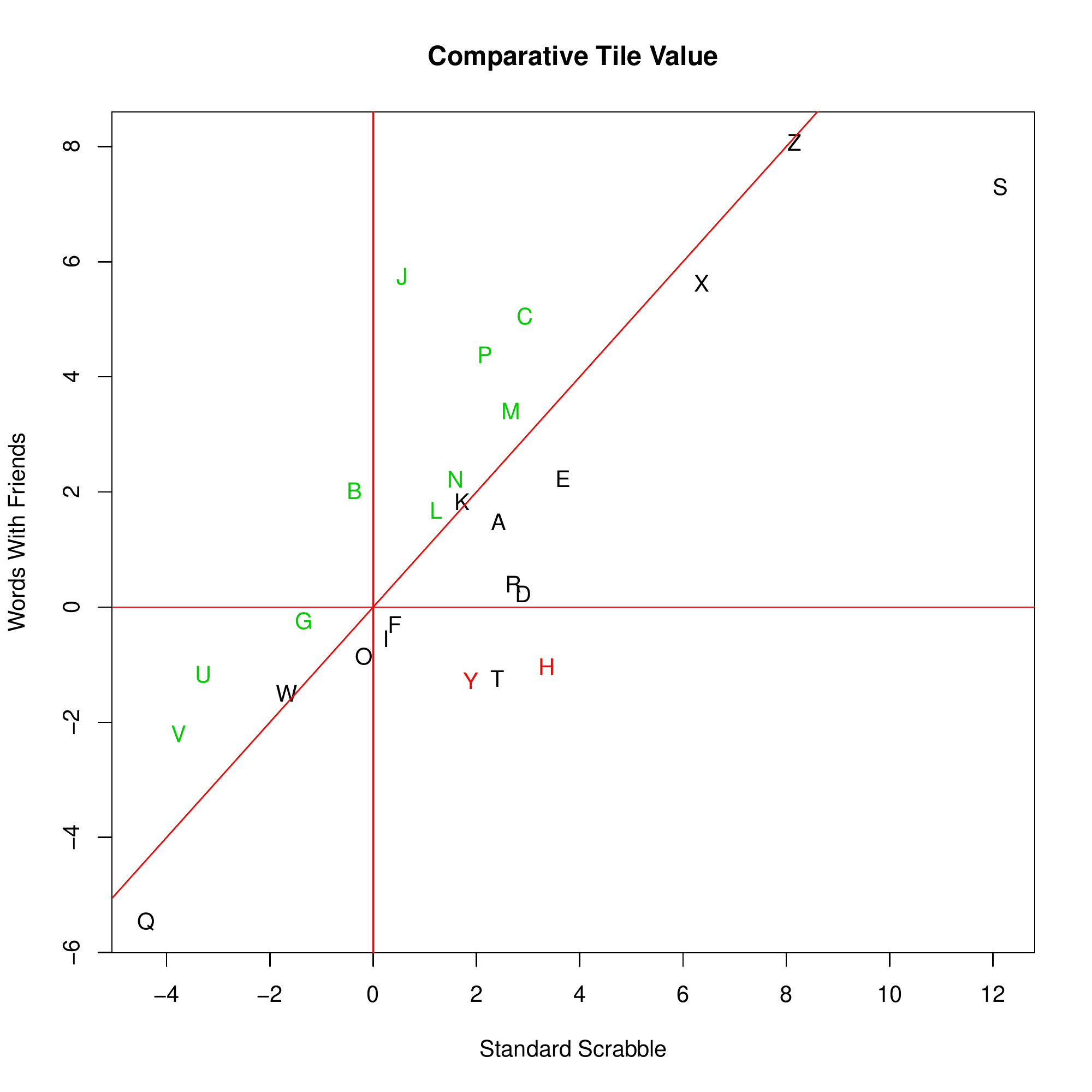}
\caption{For each letter in the bag, comparing the average point swing for Player 1 relative to the likely number of each letter the player would have access to, for each of Scrabble and Words With Friends. Letters in green have a higher value in Words With Friends than in Scrabble; those in red have lower values. The blank tile is at coordinates (31.5, 24.4). \label{fig:letter-scatter}}
\end{figure}


More generally, there is the issue of the value of any type of tile in the bag. The simplest way to estimate this value is to take the number of tiles played in a game and count the number of each particular tile that is observed, and fit a linear model to the player 1 score against the tile count. While this is imperfect, as since a player's tile exchange may interfere with draws from the middle of the sequence, the approximation is reasonable for demonstration purposes. Figure \ref{fig:letter-scatter} shows the relative value of each type of tile in the alphabet.

Along with the two blanks and four/five S'es, the four tiles with point value 8 or 10, the X, Z, J and Q, are known as the ``power tiles'' for their reputation of being beneficial to the player who draws them. This is certainly not the case for the Q; as the first player is less likely to have the Q in their rack, their mean score goes up, an average difference of roughly 5 points. Even with the relatively high score of words containing the Q, the main consequence is that the ability to play bingos is compromised by its presence. It is also apparent that the U, V and W are markedly undervalued as well, with a cost of 2 to 4 points for each, even with an upgrade in Words With Friends.

The impact of the J is fairly neutral in Scrabble, but a six-point boon in Words With Friends, as that tile has been upgraded to ten points from eight. The X and Z both yield positive benefits to the player who controls them, to the tune of 6 and 8 points each, in either version of the game.

\section{Conclusions and Proposals}

The two-sided draw scheme is an idea that has considerable theoretical advantages, but is not too likely to be used for real tournament matches short of a major technological breakthrough. Systems have been designed for the aforementioned Duplicate Bridge, but since it's much easier for a machine to deal cards than rearrange tiles, much quicker for a human to pre-deal a set of hands than re-arrange 100-odd tiles, and a much smaller market for it than for the tournament world of Duplicate Bridge.

This leaves us with the notion that we can tinker with the values of each tile in order to balance the game. Additional simulations on the Words With Friends suggest that the value of the J increases by 4 points for every 2-point change, and a similar rule might hold for other tiles. At this level it is clear that recipients of blanks and S'es benefit greatly, and subsequent reductions of the values of these tiles (perhaps even penalties for playing them!) would decrease the additional variance caused by tile order changes. 

Nonetheless, we cannot overinterpret these results immediately. Because the results of these simulations come from non-human players with their own idiosyncracies, it is not my intention to claim that the results for tile values would necessarily be duplicated on true human tournament testing, especially for players whose ability to find ``bingo'' plays is considerably lesser than the top human players; indeed, the Q and J would be more valuable in such cases since they would not be detracting from the opportunity to play bingos. However, it is clear that the substantial reduction in variance given by the duplicate format would be of great use in assessing player ability, whether or not this is due to an adjustment in a true duplicate setting, or a shift made by calculating a standard difference using simulations in this fashion.

There is hope that the same ideas can be used to get more fine-level detail for value within a game. The tiles present on the board clearly affect the value of a tile -- a triple letter score open to the left of an I greatly enhances the value of an X or a Q, for example. Taking game situation into account has been considered by \citet{shirley2010smfs}, and there is every reason to believe that the two-sided draw can be combined with this to get finer-scale assessments of player ability with fewer games played.

\subsubsection*{Acknowledgements}

I thank Mark Glickman for useful discussions on the subject of paired comparisons, and for introducing me to the work of Kenny Shirley. I also thank Mark Anderson of{\em IEEE Spectrum} for our discussions and his suggestion to compare Scrabble to Words With Friends.

\bibliographystyle{\baseloc ims}
\bibliography{\baseloc actbib}



\end{document}